\documentstyle[prl,aps]{revtex}
\begin{document}
\preprint{UCF-CM-95-001}
\title
{Ensemble density functional theory of the fractional quantum Hall effect}
\author{O. Heinonen, M.I. Lubin and M.D. Johnson}
\address{
Department of Physics, University of Central Florida, Orlando, FL 32816-2385
}
\maketitle
\begin{abstract}
We develop an ensemble density functional theory for the fractional
quantum Hall effect using a local density approximation. Model
calculations for edge reconstructions of a spin-polarized quantum
dot give results in good agreement with semiclassical and Hartree-Fock
calculations, and with small system numerical diagonalizations. This
establishes the usefulness of density functional theory to study the
fractional quantum Hall effect, which opens up the possibility of
studying inhomegeneous systems with many more electrons than has
heretofore been possible.
(Contact mdj@physics.ucf.edu.  Cond-mat paper cond-mat/9506141.)
\end{abstract}

The fractional quantum Hall effect (FQHE) is manifested in a two-dimensional
electron gas (2DEG) in a strong magnetic
field perpendicular to the plane of the
electrons\cite{Girvin}.
The effect is due to the electron-electron interactions,
which cause a downward cusp in ground state energy as a function of filling
factor $\nu=2\pi\ell_B^2 n$ at certain rational fillings $\nu=p/q$.
Here $\ell_B=\sqrt{\hbar c/(eB)}$ is the magnetic
length, $B$ the magnetic field strength, and $n$ the electron
density. These
downward cusps give rise to an energy gap, and so for these values
of $\nu$, the ground state of the system is
an incompressible liquid.
Experimentally,
FQHE systems are accessible in strip geometries, Corobino geometries,
and more recently in quantum dots\cite{McEuen},
with as few as $\approx50$ electrons, as well as
wide-well
heterojunctions and double layer systems. Theoretically, different aspects of
FQHE systems can be modeled by Laughlin's wavefunction\cite{Laughlin},
by Hartree-Fock\cite{MacD_Yang,Chamon} or composite Fermion
Hartree\cite{Brey,Chklovskii2}; by semiclassical methods
\cite{Beenakker,Chklovskii,Ferconi},
by field
theoretical approaches\cite{Chern}, which also exist for the
edge excitations on the boundary
of FQHE systems\cite{Wen}, and by exact numerical diagonalizations.
At the present, numerical diagonalizations are limited to systems with
of the order of 10 electrons. It is highly desirable to have a
computational approach which accurately treats inhomogeneous systems
with the order of
$10^2$--$10^3$ electrons.
One such approach which is in principle valid
for any interacting electron system is the
density functional theory (DFT)\cite{Kohn_Vashista,Dreizler,Parr}.

We have developed an ensemble DFT scheme within the local density approximation
(LDA)
for the fractional quantum Hall effect, and applied it to spin-polarized
circularly
symmetric quantum dots\cite{generalize}.
The results are in good agreement with results
obtained by semiclassical\cite{Beenakker,Chklovskii,Ferconi},
Hartree-Fock\cite{MacD_Yang,Chamon} (for cases where the
correlations do not play a major role), or exact diagonalization
methods\cite{Johnson}.
Our calculations show that the exchange and
correlation effects are very well represented by the LDA and that
our approach provides a computational scheme to model large inhomogeneous
FQHE systems. We note that
there exist previous formal DFTs
for strongly correlated systems, in particular for high-temperature
superconductors \cite{Gross}, and DFT calculations of high-T$_{\rm c}$
materials \cite{Pickett} and transition-metal oxides \cite{Svane}.
Ferconi, Geller and Vignale\cite{Ferconi} have also recently
studied FQHE systems within the DFT
using an extended Thomas-Fermi
approximation, including a LDA for the exchange-correlation
energy.
However, ours are, to the
best of our knowledge, the first practical DFT-LDA calculations of
a strongly correlated system in strong magnetic fields, and demonstrate the
usefulness of the DFT-LDA in studying large inhomogeneous FQHE systems.

In typical DFT calculations of systems of $N_{\rm el}$ electrons,
the standard Kohn-Sham (KS) scheme\cite{KohnSham} is implemented,
in which the particle density $n({\bf r})$ is expressed in terms of
a Slater determinant of $N\geq N_{\rm el}$
KS orbitals, $\psi_{\alpha}({\bf r})$. These obey an effective
single-particle Schr\"odinger equation
$H_{\rm eff}\psi_\alpha=\epsilon_\alpha\psi_\alpha$, which is solved
self-consistently by occupying the $N_{\rm el}$ KS orbitals with the lowest
eigenvalues $\epsilon_\alpha$ (we identify the Fermi energy of the
system with the largest
$\epsilon_\alpha$ of the occupied orbitals),
and iterating.
This scheme works well in practice
for systems which are noninteracting $v$-representable\cite{Dreizler,Parr},
{\em i.e.,\/}
systems for which the true
electron density can be represented by a single Slater
determinant of single-particle wavefunctions.
However, when the KS orbitals are degenerate at the Fermi energy
there is an
ambiguity in how to occupy these degenerate orbitals.
This is the case for the FQHE, as we now demonstrate.
Consider an FQHE system in
the $xy$-plane with the magnetic field along the $\hat z$-axis.
A circularly symmetric external potential
$V_{\rm ext}({\bf r})=V_{\rm ext}(r)$
(due, {\em e.g.,\/} to a uniform positive background charge density)
confines the systems such that the density
is fixed with a local filling of $\nu=1/3$ up to an edge at $r_0$
($r_0\gg\ell_B$)
where the density falls to zero within a distance of order $\ell_B$.
That such
systems exist is well demonstrated by the excellent agreement
between the Laughlin wavefunction and experiments, and by many
numerical
calculations\cite{Haldane,Johnson}.
Due to the circular symmetry
we can label
single-particle orbitals by
angular momentum $m$, and by a ``band'' or Landau level index $n\geq0$.
The orbitals $\psi_{m,n}({\bf r})$ are centered
on circles of radii $r_m\approx\sqrt{2m}\ell_B$ with Gaussian fall-offs
for $r\ll r_m$ and $r\gg r_m$.
The single-particle orbitals with $n=0$ are then in the bulk all degenerate,
and the
degeneracy is not lifted by electron-electron interactions since the system is
homogeneous in the bulk.
In order to obtain a constant density at $\nu=1/3$ even at the center of the
system,
all single-particle orbitals  in the bulk with $n=0$ must
have occupancies $1/3$. If the Fermi energy lay above the energies of the
bulk orbitals, they would all be filled and one would have $\nu=1$.
Therefore, to get occupancies $1/3$ the Fermi energy must lie at
the degenerate energy $\epsilon_{m0}$
of these orbitals.
Thus,
in applying DFT to the FQHE we can expect a huge degeneracy of KS orbitals
at the Fermi energy. Consequently, the particle density cannot be expressed
in terms of a single Slater determinant. Instead,
the density has to be constructed from an ensemble of Slater determinants,
{\em i.e.,\/} the orbitals at the Fermi energy are assigned fractional
occupation numbers, just as
the Laughlin wavefunction is not a single Slater determinant,
but a highly correlated state with average occupancies of 1/3 of
single-particle states.
This is generally known as
{\em ensemble} density functional theory\cite{Dreizler,Parr}.

Although ensemble DFT has been developed formally, there are in practice
few examples of applications and calculations using ensemble DFT
for ground state calculations. A significant aspect of our
work is that we have developed an ensemble scheme which is practical
and useful
for the study of the FQHE.
In ensemble DFT, any physical density
$n({\bf r})$ can be represented by
$
n({\bf r})=\sum_{mn}f_{mn}|\psi_{mn}({\bf r})|^2,
$
where $f_{mn}$ are occupation numbers satisfying
$0\leq f_{mn}\leq1$, and the orbitals $\psi_{mn}$
satisfy the equation
\begin{equation}
\left\{
\frac{1}{2m^*}\left[{\bf p}+\frac{e}{c}{\bf A}({\bf r})\right]^2
+V_{\rm ext}({\bf r})+V_{\rm H}({\bf r})
+V_{\rm xc}({\bf r},{\bf B})\right\}\psi_{(m,n)}({\bf r})=\epsilon_{mn}
\psi_{mn}({\bf r}),\label{HK}
\end{equation}
where $\nabla\times{\bf A}({\bf r})={\bf B}({\bf r})$.
In equation (\ref{HK}),
$V_{\rm H}({\bf r})$ is the
Hartree interaction of the 2D electrons, and
$V_{\rm xc}({\bf r},{\bf B})$ is the exchange-correlation potential, defined as
a functional derivative of the exchange-correlation energy
$E_{\rm xc}[n({\bf r}),{\bf B}]$
of the system with respect to density:
$
V_{\rm xc}({\bf r},{\bf B})=\left.{\delta E_{\rm xc}[n({\bf r}),{\bf B}]
\over\delta n({\bf r})}\right|_{\bf B}.
$
We will hereafter not explicitly indicate the parametric dependence
of $V_{\rm xc}$ and
$E_{\rm xc}$ on $\bf B$.
The question is then how to determine
these orbitals and their occupancies in the presence of degeneracies. Here, we
have devised a scheme to obtain a set of occupancies which (a) converges
to physical densities (to the best of our knowledge) for FQHE systems,
and (b) reproduces finite temperature DFT as well as the standard KS
scheme for noninteracting $v$-representable systems.
This scheme may be of much more general applicability to general systems
which are not noninteracting $v$-representable
other than the FQHE. In our
scheme, we start with input occupancies and single-particle orbitals and
iterate the system $N_{\rm eq}$ times using the KS scheme.
The number $N_{\rm eq}$ is chosen large enough (about 20--30 in practical
calculations)
that the density is close to the final density after $N_{\rm eq}$ iterations.
Were the system noninteracting $v$-representable, we would now essentially
be done. However, in this system there are now in general many degenerate
or near-degenerate orbitals at the Fermi energy, and small fluctuations in the
density between iterations cause the KS scheme to occupy a different
subset of these orbitals each iteration. This corresponds to constructing
different Slater determinants each iteration. While the occupation numbers
$f_{mn}$ of these orbital are zero or unity more or less at random
each iteration, the {\em average} occupancies, {\em i.e.,\/} the
occupancies averaged over many iterations, become well defined and approach the
value, say, 1/3 for orbitals localized
in a region where the local filling factor is close to $\nu=1/3$.
We use this to construct an ensemble by
accumulating running average occupancies $f_{mn}$
after the initial $N_{\rm eq}$ iterations
and use these to calculate densities.
Thus, our algorithm essentially
picks a different (near-)degenerate Slater determinant after each iteration,
and these
determinants are all weighted equally in the ensemble.
It is clear that this scheme reproduces
the results of the KS scheme for noninteracting $v$-representable systems
(for which the KS scheme picks only the one Slater determinant which gives
the ground state density)
for $N_{\rm eq}$ large enough. We have numerically
verified that a finite-temperature version of our scheme converges
to a thermal ensemble at finite temperatures down to
temperatures of the order of $10^{-3}\hbar\omega_c/k_B$. We
have also performed preliminary Monte Carlo simulations about the
ensemble obtained by our scheme. The results are that to within numerical
accuracy our scheme gives the lowest energy.

In the LDA, the exchange-correlation
energy is assumed to be a local function of density,
$
E_{\rm xc}/N=\int d{\bf r}\epsilon_{\rm xc}(\nu)n({\bf r}),
$
where $\epsilon_{\rm xc}(\nu)$ is the exchange-correlation energy per
particle in a {\em homogeneous} system of constant density
$n=\nu/(2\pi\ell_B^2)$ and
filling factor $\nu$.
Experience has shown that the LDA often works surprisingly well,
even
for systems in which the electron density is strongly
inhomogeneous\cite{Kohn_Vashista}. In the
FQHE, the length scale of exchange-correlation interactions
and density fluctuations is
given by the magnetic length $\ell_B$ due to the Gaussian fall-off of
any single-particle basis in which the interacting Hamiltonian
is expanded. The densities are relatively smooth on this length scale,
which gives us additional hope that the LDA will work well for the FQHE, too.
For the exchange-correlation energy per
particle of
a uniform electron gas in a constant magnetic field we use the Pad\'e
approximant\cite{Rasolt}
$
\epsilon_{\rm xc}(\nu)={\epsilon_{\rm xc}^{\rm L}(\nu)+
\nu^4\epsilon_{\rm xc}^{\rm TC}(n(\nu))\over 1+\nu^4},
$
where $\epsilon_{\rm xc}^{\rm TC}$ is the zero-magnetic field result
\cite{Tanatar}.
The term $\epsilon_{\rm xc}^{\rm L}(\nu)$
consists of two terms. The first one is a smooth interpolation
formula\cite{Levesque} $\epsilon_{\rm xc}^{\rm LWM}(\nu)$
between ground state
energies at some rational fillings. The
second one, $\epsilon_{\rm xc}^{\rm C}(\nu)$,
is all-important for the study of the FQHE. This term contains the
cusps in the ground state energy which cause the FQHE. Here we have used
a simple model which captures the essential physics. We model
$\epsilon_{\rm xc}^{\rm C}(\nu)$ by constructing it to be zero at
values of $\nu=p/q$ which display the FQHE. Near $\nu=p/q$,
$\epsilon_{\rm xc}^{\rm C}(\nu)$ is linear and
has at $\nu=p/q$ a discontinuity in the
slope related to the chemical potential gap
$\Delta \mu=q(|\Delta_p|+|\Delta_h|)$. Here $\Delta_{p,h}$ are the
quasiparticle (hole) creation energies  which
can be obtained from the literature \cite{Morf_Halperin,Morf_Ambrumenil}
at fractions $\nu=p/q$.
Farther
away from $\nu=p/q$, $\epsilon_{\rm xc}^{\rm C}(\nu)$ decays to zero.
Finally, in
the LDA $V_{\rm xc}(r)$ is obtained from $\epsilon_{\rm xc}(\nu)$ as
$
V_{\rm xc}(r)=\left.{\partial \left[\nu\epsilon_{\rm xc}(\nu)\right]
\over\partial\nu}\right|_{\nu=\nu(r)}
$
at constant $B$. In our calculations, we restrict ourselves to include
only the cusps at $\nu=1/3,2/5,3/5$ and $\nu=2/3$, which are the
strongest fractions.

A technical difficulty arises in the LDA: The
discontinuities in $V_{\rm xc}(r)$ in the LDA give rise to a
numerical instability. The reason
is that an arbitrarily small fluctuation in charge density close to an FQHE
fraction gives rise to a finite change in energy. To overcome this
problem, we made the compressibility of the system finite, but very small,
corresponding to a finite, but very large, curvature instead of a point-like
cusp in $\epsilon_{\rm xc}$ at the FQHE fractions. What we found worked very
well in practice was to have the discontinuity in chemical potential occur
over an interval of filling factor of magnitude $10^{-3}$. This corresponds
to a sound velocity of about $10^6$ m/s in the electron gas, which is
three orders of magnitude larger than the Fermi velocity of a 2D electron
gas at densities typical for the FQHE.
Figure \ref{V_xc} depicts a plot of $V_{\rm xc}$ as a function of
filling factor used in our calculations.

We have self-consistently solved the KS equations Eqs. (\ref{HK})
for a spin-polarized quantum dot in a parabolic external potential,
$V_{\rm ext}(r)=\frac{1}{2}m^*\Omega^2r^2$ by expanding the KS
orbitals $\psi_{mn}({\bf r})=e^{im\phi}\varphi_{mn}(r)$ in the eigenstates
of $H_0=\frac{1}{2m^*}\left({\bf p}-\frac{e}{c}{\bf A}({\bf r})\right)^2$
in the cylindrical gauge, ${\bf A}({\bf r})=\frac{1}{2}Br\hat\phi$,
including the four lowest Landau levels ($n=0,\ldots,3$). We
chose $\epsilon_0=13.6$, appropriate for
GaAs, and a confining potential of strength\cite{McEuen}
$\hbar\Omega=1.6$ meV.
In particular, we have used our DFT-LDA scheme to study the
edge reconstruction of the quantum
dot as a function of magnetic field strength. As is known from
Hartree-Fock and exact
diagonalizations\cite{MacD_Yang,Chamon,Johnson,Brey,Chklovskii2},
for strong confinement the
quantum dot forms a maximum density droplet
in which the density is uniform
at $\nu=1$ in the interior, and falls off rapidly to zero at
$r\approx\sqrt{2N}\ell_B=r_0$.
As the magnetic field strength increases,
a ``lump'' of density breaks off, leaving a ``hole'' or deficit at about
$r=r_0$. This effect is due to the short-ranged attractive
exchange interaction:
it is energetically favorable to have a lump of density break off so that the
system can take advantage of the exchange energy in the lump. As $B$ is
further increased, the correlations will cause
incompressible strips with densities $\nu=p/q$ to appear
\cite{Beenakker,Chklovskii,Gelfand,Ferconi} on the edges,
and incompressible droplets to form in the bulk
at densities $\nu=p/q$.
Figure \ref{reconstr} depicts various stages
of edge
reconstruction obtained by us as the magnetic field strength is increased.
The value of
$\Omega$
for which the exchange lump appears compares very well with the value
found by De Chamon and Wen\cite{Chamon}
in Hartree-Fock and numerical diagonalizations.
At higher fields still, the incompressible strips appear
at the edges, and incompressible droplets are formed in the bulk.
We emphasize that incompressible regions that appear in our calculations
are not due to the presence of a
uniform positive background density which tends to fix the bulk density at
the value of the background density.

There is also another edge effect caused by
correlations. For particular, stiff confining potentials, so-called
composite edges\cite{MacDonald,Johnson}
can appear. These can be thought of as particle-hole
conjugates of uniform incompressible droplets.
Consider a droplet with a bulk density corresponding to $\nu=1/3$,
falling off to zero at the
edge. A incompressible droplet with a bulk density of $\nu=2/3$ is
obtained by particle-hole conjugation.
However, at the edge, the density will
first {\em rise} to $\nu=1$ (since the density of the $\nu=1/3$ droplet
drops to zero), and then eventually drop to zero. Note that this
argument is based on particle-hole conjugation, which is
an exact symmetry of the lowest Landau level\cite{symmetry}, and it is unclear
if composite edges exist in real systems, which do not strictly obey
particle-hole symmetry.

Figure \ref{composite} depicts the occupations of the Kohn-Sham orbitals and
the
particle density (inset) for a system where the confining potential is
supplied by a uniform positive background charge density $n_+=2/(6\pi\ell_B^2)$
for $r<r_0$, where
$r_0$ is fixed by charge neutrality.
{}From this figure, we see that for this choice of potential,
the system forms a composite edge,  even though our system does
not obey particle-hole symmetry. We therefore conclude that such
edges can exist in real systems. We have also verified the stability of
all our incompressible regions by adding or subtracting a particle
from the system.

The authors would like to thank M. Ferconi, M. Geller and G. Vignale for
helpful
discussions and for sharing their results prior to publications, and
K. Burke and E.K.U Gross for useful comments about the DFT. O.H. would like to
thank Chalmers Institute of Technology, where part of the numerical work was
done.
This work was supported by
the NSF through grant DMR93-01433.

\begin{figure}
\caption{Exchange-correlation potential $V_{\rm xc}$ as function of
filling factor in units
of $e^2/(\epsilon_0\ell_B)$ for $0\leq\nu\leq1$.
The increase in $V_{\rm xc}$
at an FQHE filling occurs over a range of filling factor of 0.004.}
\label{V_xc}
\end{figure}
\begin{figure}
\caption{Edge reconstruction of a quantum dot as the magnetic field
strength is increased.
Plotted here is the local filling factor
$\nu(r)$ for a parabolic quantum dot with
$\hbar\Omega=1.6$ meV and
40 electrons.
For magnetic field strengths $B< 2.5$ T the dot forms a maximum density
droplet, and for $B\approx3.0$ T, an exchange hole is formed. For stronger
magnetic fields, incompressible regions form, separated by compressible
strips.}
\label{reconstr}
\end{figure}
\begin{figure}
\caption{Occupancies $f_{m0}$ {\em vs} orbital centers
$r_m$ for a composite edge of
a system of 45 electrons (diamonds). Here $B=5.0$ T.
Near the edge, the occupancies rise to unity. The full line shows the
corresponding local filling factor.}
\label{composite}
\end{figure}

\begin{references}
\bibitem{generalize} Generalizations to include spin degrees of freedom will
be presented in a future publication.
\bibitem{symmetry} Note, however, that for a finite system
the Hilbert space for the holes
has to be restricted. See Ref. \onlinecite{Johnson}.
\end{references}
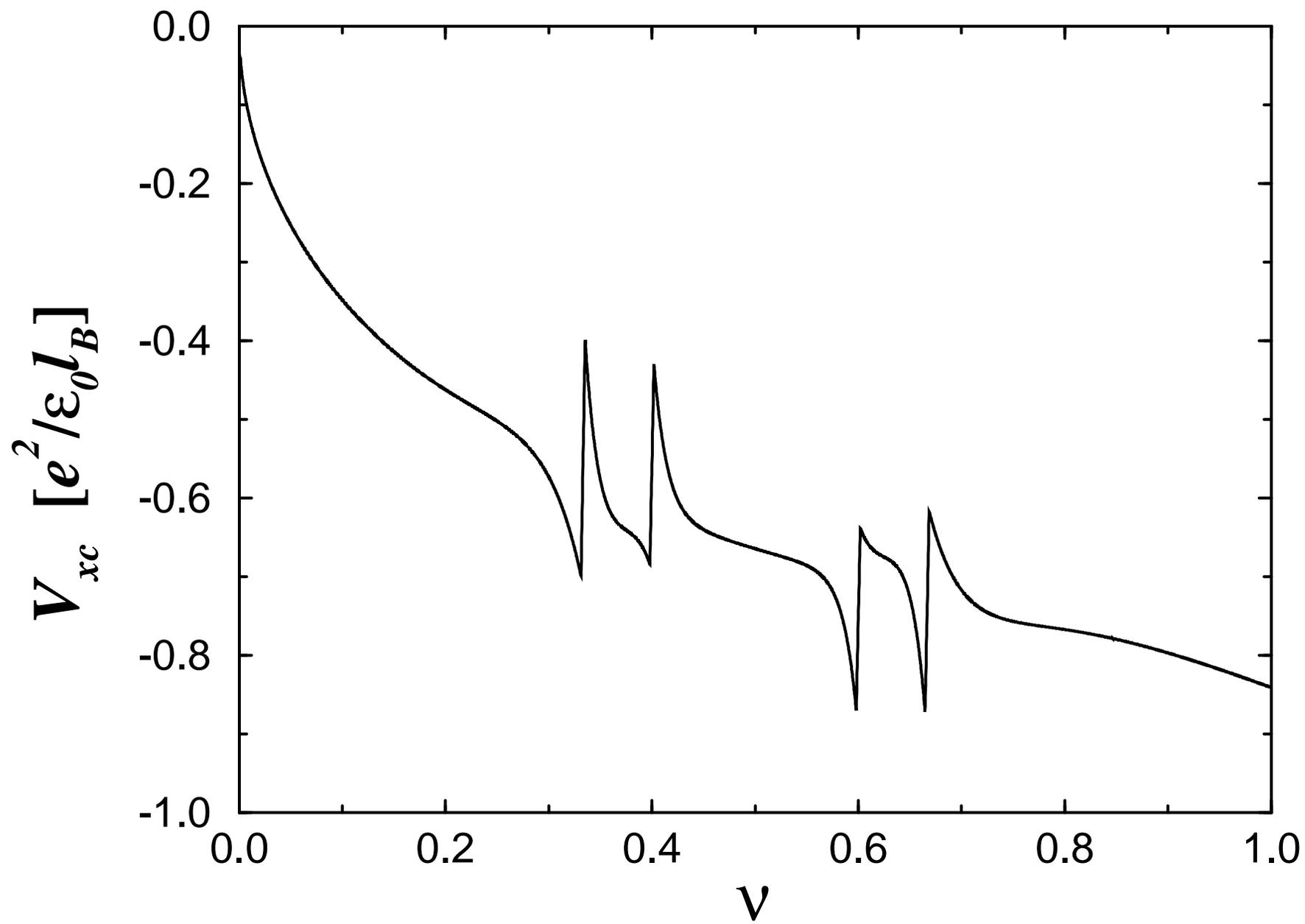
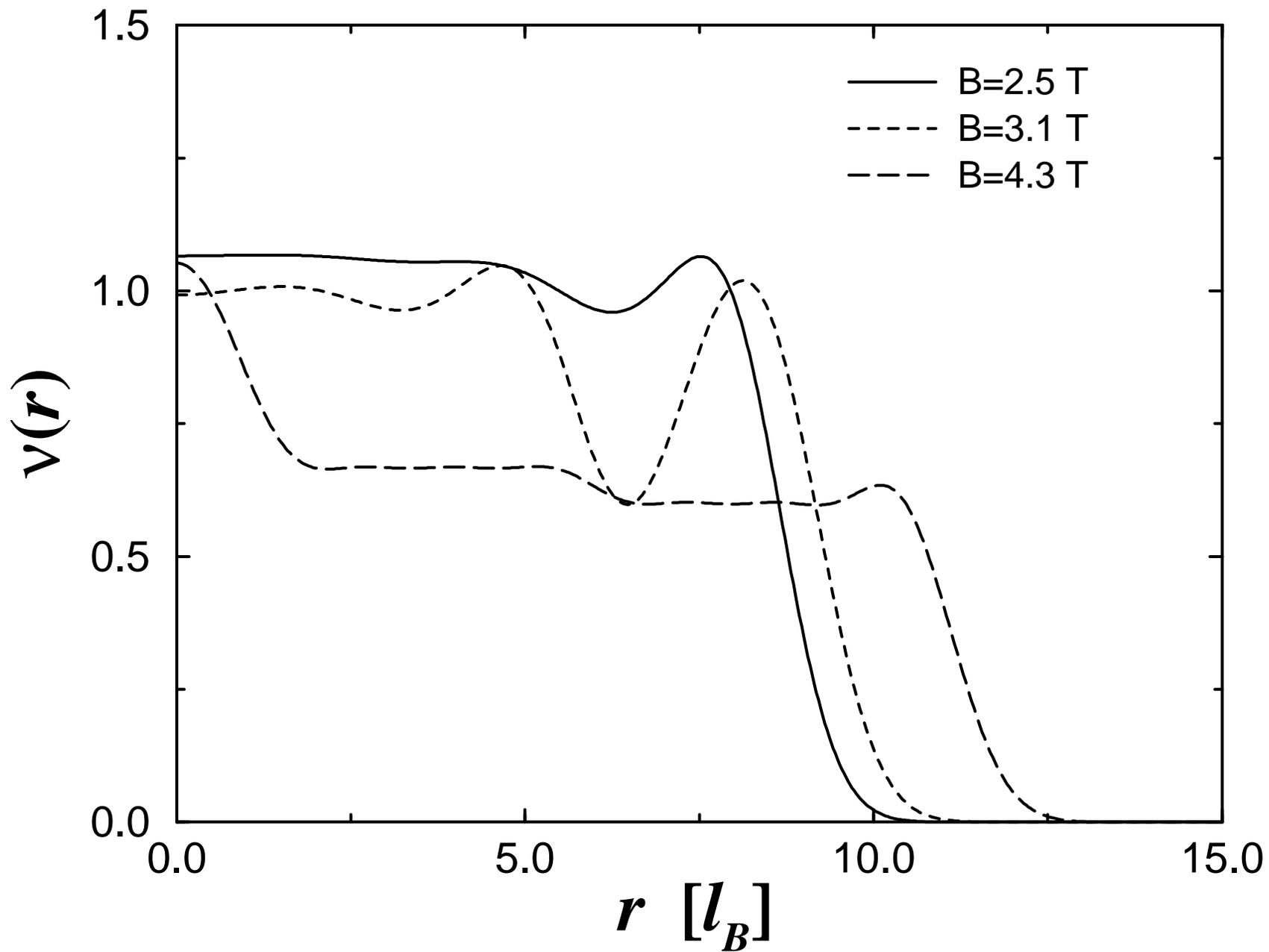
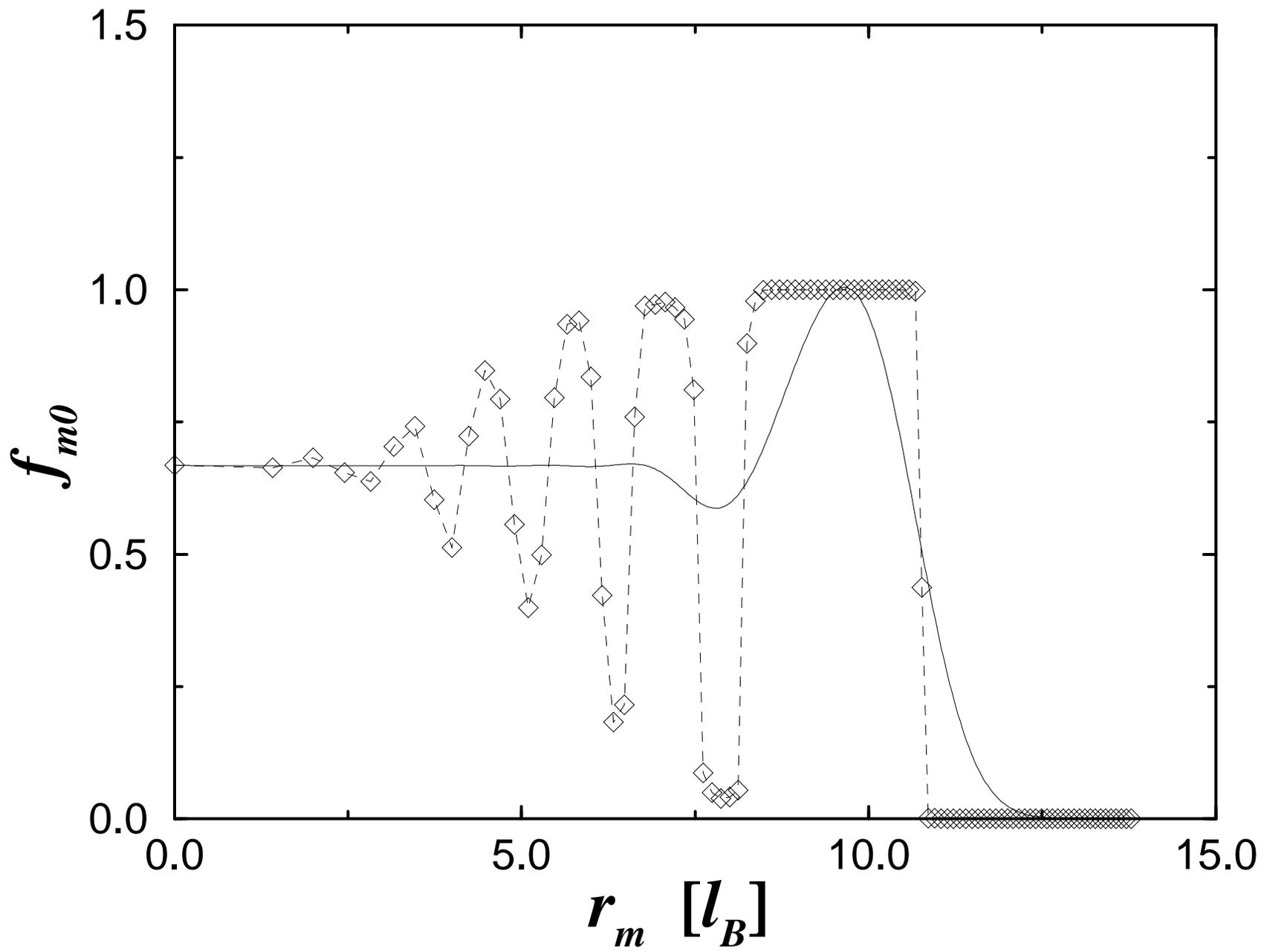
\end{document}